\documentclass[twocolumn,amssymb,amsfonts,showkeys, reprint, nobibnotes, aps, pra]{revtex4-2}
\usepackage{bm}
\usepackage[T1]{fontenc}
\usepackage{amssymb}
\usepackage{MnSymbol}
\usepackage{amsthm}
\usepackage{mathtools}
\usepackage{physics}
\usepackage{xcolor}
\usepackage{graphicx}
\usepackage[left=23mm,right=13mm,top=35mm,columnsep=15pt]{geometry} 
\usepackage{adjustbox}
\usepackage{placeins}
\usepackage[T1]{fontenc}
\usepackage{lipsum}
\usepackage{csquotes}

\DeclareUnicodeCharacter{2212}{-}
\begin{document}
	\title{Anomalous Lasing Behavior in a Nonlinear Plasmonic Random Laser}
\author{Renu Yadav$^1$, Sourabh Pal$^2$, Subhajit Jana$^1$, Samit K. Ray$^1$ Maruthi M. Brundavanam$^1$, and Shivakiran Bhaktha B. N. $^1$} 	\email[Correspondence email address: ]{kiranbhaktha@phy.iitkgp.ac.in}
\affiliation{$^1$Department of Physics, Indian Institute of Technology Kharagpur, Kharagpur-721302, India \\ $^2$Advanced Technology Development Centre, Indian Institute of Technology Kharagpur, Kharagpur 721302, India}
	\begin{abstract}
An unprecedented double-threshold lasing behavior has been observed in a plasmonic random laser composed of Au nanoislands decorated on vertically standing ZnO nanorods, infiltrated with dye-doped polymer matrix. The strong coupling of random laser modes to plasmonic nanocavities results in a dominant absorption of the random laser emission, leading to the first unusual lasing threshold. At higher pump fluences, the nonlinear optical behavior of the Au nanoislands induces a second lasing threshold. Various statistical tools have been employed to analyze the intensity fluctuations of the random laser modes, validating this unique lasing behavior.
	\end{abstract}
 
	
	\maketitle
	 \section{Introduction}
	Random lasers (RLs) are mirror-less optical devices that employ scattering from the disordered medium for the optical feedback. The simpler and cost-effective fabrication of RLs, made possible by the absence of well-defined resonant cavities, comes at the expense of control over emission wavelength and direction \cite{cao1999random,lawandy1994laser}. Among the various RLs realized using different scattering media \cite{tulek2010studies,lu2015random,cao1999random,song2010random,lawandy1994laser,shivakiran2012optofluidic,ferjani2008random}, plasmonic random lasers (PRLs) employing metal nanoparticles (MNPs) for scattering feedback have garnered significant attention. This interest is driven by their superior absorption and scattering cross-sections compared to dielectric nanoparticles (NPs) of similar dimensions, as well as the local field enhancement provided by localized surface plasmon resonance (LSPR) \cite{maier2007plasmonics}. In addition to their excellent scattering properties, MNPs can alter the decay and excitation rates of fluorophore molecules (gain media), influenced by their spatial separation and the spectral overlap of their emission and absorption spectra \cite{kummerlen1993enhanced,lakowicz2006plasmonics}. MNPs offer strong scattering feedback and improved fluorophore excitation and emission efficiency, thereby lowering the lasing threshold in highly open and lossy random lasing systems \cite{dice2005plasmonically,popov2006random,meng2011plasmonically,meng2013metal,Dominguez:11,chandrasekar2017lasing,zhai2011random,cao2012plasmon,ziegler2015gold,yin2016shape}.

In RLs, high pumping rates can induce nonlinear effects resulting from gain saturation and mode competition \cite{andreasen2011modes,andreasen2011nonlinear}. The strong local electromagnetic fields in plasmonic NPs can boost these nonlinear optical effects in PRLs \cite{kauranen2012nonlinear}. Among the various  plasmonic NPs, silver (Ag) and gold (Au) NPs have been widely used as their LSPR peaks lies in visible-near infrared region \cite{maier2007plasmonics,loiseau2019core}. Ag has low absorption losses and high tunability as compared to Au NPs, however Ag has low stability and is toxic for biological systems. The higher stability, biocompatibility and chemical inertness of Au NPs makes them more suitable for various applications \cite{rycenga2011controlling,marin2015applications,boisselier2009gold,dreaden2012golden}. Optical nonlinear behaviour in Au NPs of different shapes and sizes has been explored in detail in literature \cite{elim2006observation,olesiak2012third,ara2012diffraction,dong2011shape,philip2012evolution}.
It has also been observed that the plasmonic modes of MNPs can couple to nanophotonic modes of resonant cavities \cite{ameling2010cavity,ameling2013microcavity,garcia2024realization}. Depending on the strength of interaction between the two resonant modes, their coupling has been utilized for various applications such as, to enhance non-radiative energy transfer between donor and acceptor dye molecules strongly coupled to a metal Fabry–P\'erot cavity \cite{zhong2017energy}, 
to design dual-band absorbers with broad spectral range for photovoltaic applications \cite{hagglund2016strong} and for plasmonic hot-carrier generation and  transfer in metal-dielectric systems \cite{shan2019direct,zeng2016photoinduced,wong2021enhancing}.
\begin{figure*}[ht]
		\centering
		\includegraphics[width=12 cm]{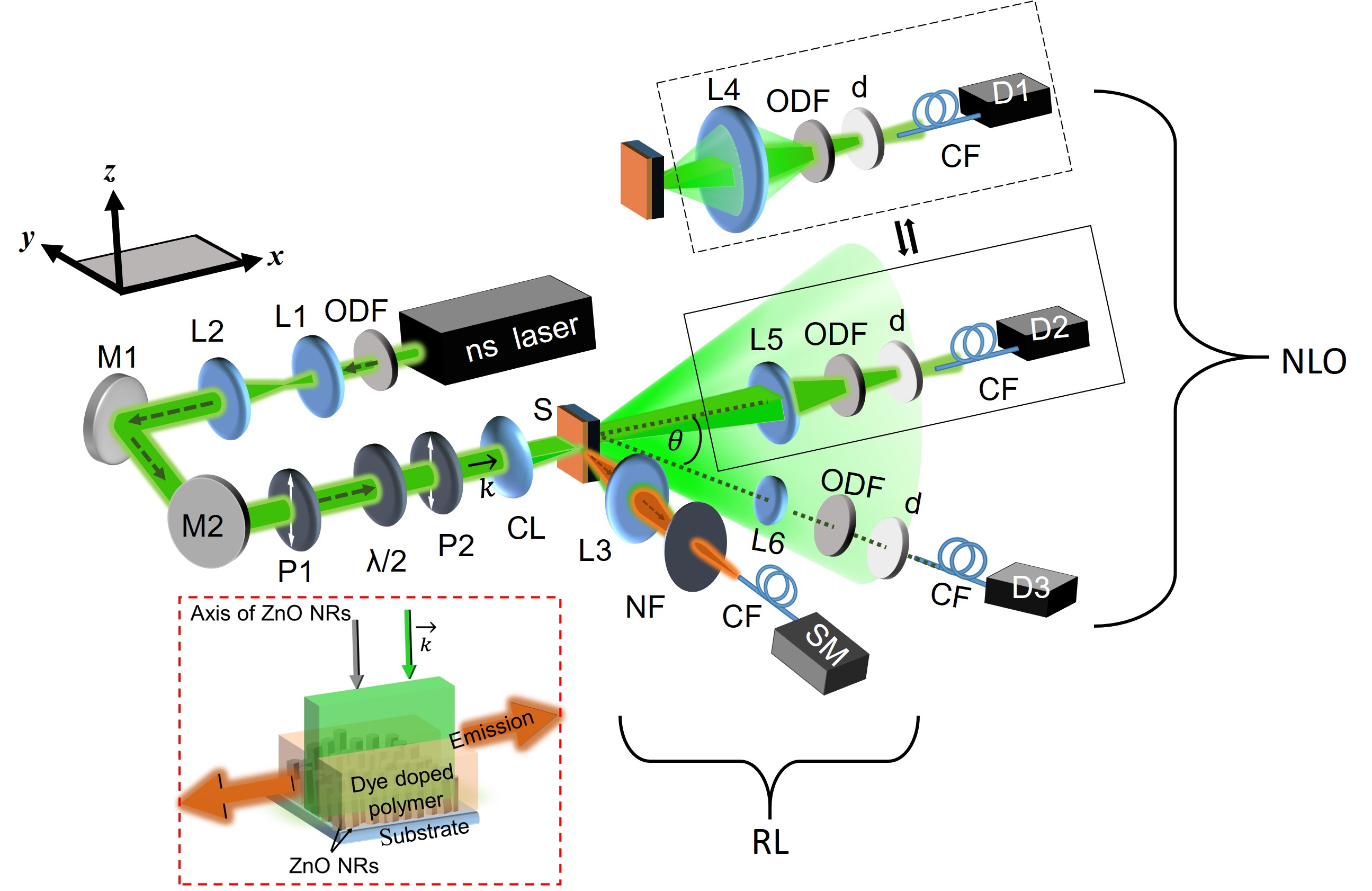}
		\caption{Schematic of the experimental setup used to study RL characteristics (RL) and nonlinear optical absorption and scattering properties (NLO) of the system. Detector D1, D2 and D3 collect total transmitted intensity which includes scattered intensity, transmitted intensity with reduced scattered intensity contribution, and intensity scattered at an angle of 5 $^\circ$ from the pump propagation direction, respectively. The inset in the red dashed rectangle shows the excitation and collection geometries in reference to the axis of ZnO NRs, $\vec{k}$. The labeled optical components are, ODF: optical density filter, L1-L5: Lens, M1, M2: mirror, P1,P2 : polarizer, $\lambda/2$: half wave plate, CL: cylindrical lens, S: sample, NF: notch filter, CF: collection fiber, d: diffuser, SM, D1, D2, D3: spectrometer.}
	\label{fig27}
      \end{figure*}
      
This work investigates random lasing in a system comprising vertically aligned ZnO nanorods (NRs) adorned with Au nanoislands (NIs) and encased in a dye-doped polymer matrix (AuZnO). The randomly arranged ZnO NRs and Au NIs provide multiple scattering feedback for the photons emitted by the dye molecules. A unique double-threshold lasing behavior is observed in the presence of Au NIs. The characteristics of the lasing modes at these two thresholds are examined. The double-threshold phenomenon arises from two distinct mechanisms: initially, the dominant absorption of dye emission by plasmonic nanocavities coupled to RL modes at low pump fluence, and subsequently, enhanced nonlinear scattering and absorption of the pump beam at higher pump fluence. The study delves into the contributions of both coherent and incoherent components to the observed lasing behavior. The statistical analysis of the intensity fluctuations in the modes of AuZnO RLs, using different statistical tools such as, L\'evy statistics, covariance analysis and replica symmetry breaking (RSB), validates the observed double-threshold lasing behavior of the system.
     
\section{Experimental Details}
\subsection{Fabrication Methods}
Vertically aligned ZnO NRs were fabricated on the glass substrate by physical and chemical growth techniques \cite{pal2021boron,yadav2023synergy}. Au NIs were deposited on ZnO NRs by thermal dewetting method \cite{jana2022synergistic,yadav2023synergy}. A Pyrromethene 570 dye-doped poly(methyl methacrylate) (PMMA) matrix was spin-coated on ZnO NRs (ZnO sample) and Au decorated ZnO NRs (AuZnO sample)  to introduce gain in the system. 
\subsection{Characterization Techniques}

The morphology of ZnO NRs and Au NIs was studied using a field-emission scanning electron microscope (FESEM, Zeiss SUPRA 40).  The optical absorption spectra were recorded using an ultraviolet−visible spectrophotometer (UV−vis, model AvaSpec-3648). The dye-doped polymer matrices were excited with a continuous wave diode pumped solid state laser of 532 nm wavelength to record their photoluminescence (PL) spectra using a CCD-based fiber-probe Avantes spectrometer with a spectral resolution of 0.2 nm. The random lasing characteristics were studied by exciting the samples with the second harmonic of a Q–switched Nd:YAG laser (NANO S120–20, Litron Lasers) with a pulse width of 10 ns, repetition rate of 20 Hz, and emission wavelength of 532 nm. A combination of a half-wave plate and a polarizer was used to control the pump energy and to excite the samples with a vertically polarized pump beam with respect to the lab frame. This polarization was chosen because the anisotropic nature of dye molecules, which behave like dipoles, results in maximum emission intensity with vertical pump polarization. The polarized pump beam was focused into a stripe on the sample using a cylindrical lens of focal length, f = 5 cm. The emission spectra were recorded in the plane of the sample along the pump stripe using a CCD-based fiber-probe spectrometer with 0.04 nm spectral resolution. The schematic of the experimental setup is shown in Fig. \ref{fig27}. Two different optical setups have been used, one for random lasing studies (labeled as RL) and other for nonlinear optical studies (labeled as NLO). The details of NLO setup will be discussed in Section \ref{nls}. The inset shows the excitation and collection geometries in reference to the axis of ZnO NRs.

       \begin{figure}[ht]
		\centering
		\includegraphics[width=8.5 cm]{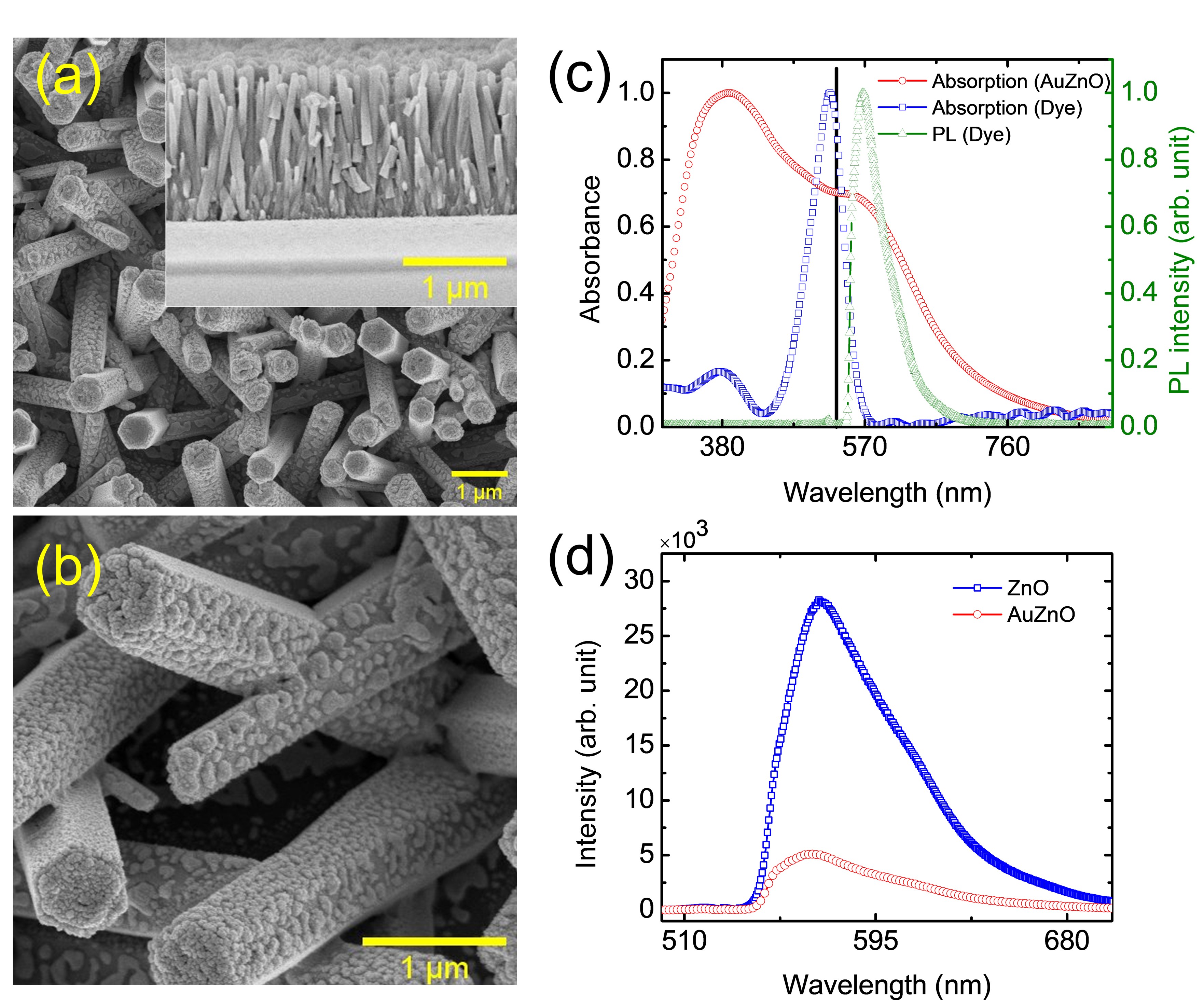}
		\caption{ (a) FESEM image of top-view of Au NI decorated ZnO NRs. The inset shows the cross sectional view of ZnO NRs. (b) A magnified view of Au NIs on ZnO NRs. (c) Extinction spectra of AuZnO, absorption and PL spectra of Pyrromethene-doped PMMA. The solid black line indicates the pump laser wavelength. (d) PL spectra of Pyrromethene-doped PMMA coated  ZnO and AuZnO samples.}
		\label{fig17}
    \end{figure}
\section{ Results and Discussion}
\subsection{Morphological and Optical Characterization}

 Fig. \ref{fig17} (a) shows the top-view of the Au NI decorated ZnO NRs. They have hexagonal cross-section with diameter in the range of $\sim$ 80 – 160 nm. The inset shows the cross-sectional image of ZnO NRs. The NRs stand vertically on glass substrate with an average height of  $\sim$ 1 $\mathrm{\mu m}$. Fig. \ref{fig17} (b) shows the magnified top-view of AuZnO NRs. The Au NIs have diameter in the range $\sim$ 20 - 150 nm. Fig. \ref{fig17} (c) shows the UV-visible extinction spectra of Au NI decorated ZnO NRs, and absorption and PL spectra of Pyrromethene doped PMMA matrix. The solid black line indicate the pump laser wavelength. A peak at $\sim 371 $ nm corresponds to  exciton absorption in ZnO \cite{haase1988photochemistry}. Another peak at $\sim$ 559 nm corresponds to LSPR due to Au NIs \cite{maier2007plasmonics}. A large dispersion in the shape, size and distribution of NIs results in a broad LSPR band. It is observed that the absorption and PL spectra of the dye have a strong spectral overlap with LSPR band which ensures a strong metal-gain interaction. Fig. \ref{fig17} (d) shows the PL emission of dye-doped polymer coated ZnO and AuZnO samples. A quenching of the fluorescence due to Au NIs is observed. This behavior can be attributed to non-radiative energy transfer from dye emitters to Au NIs \cite{yadav2023synergy}.
  \begin{figure}[ht] 
		\centering
		\includegraphics[width=8.5 cm]{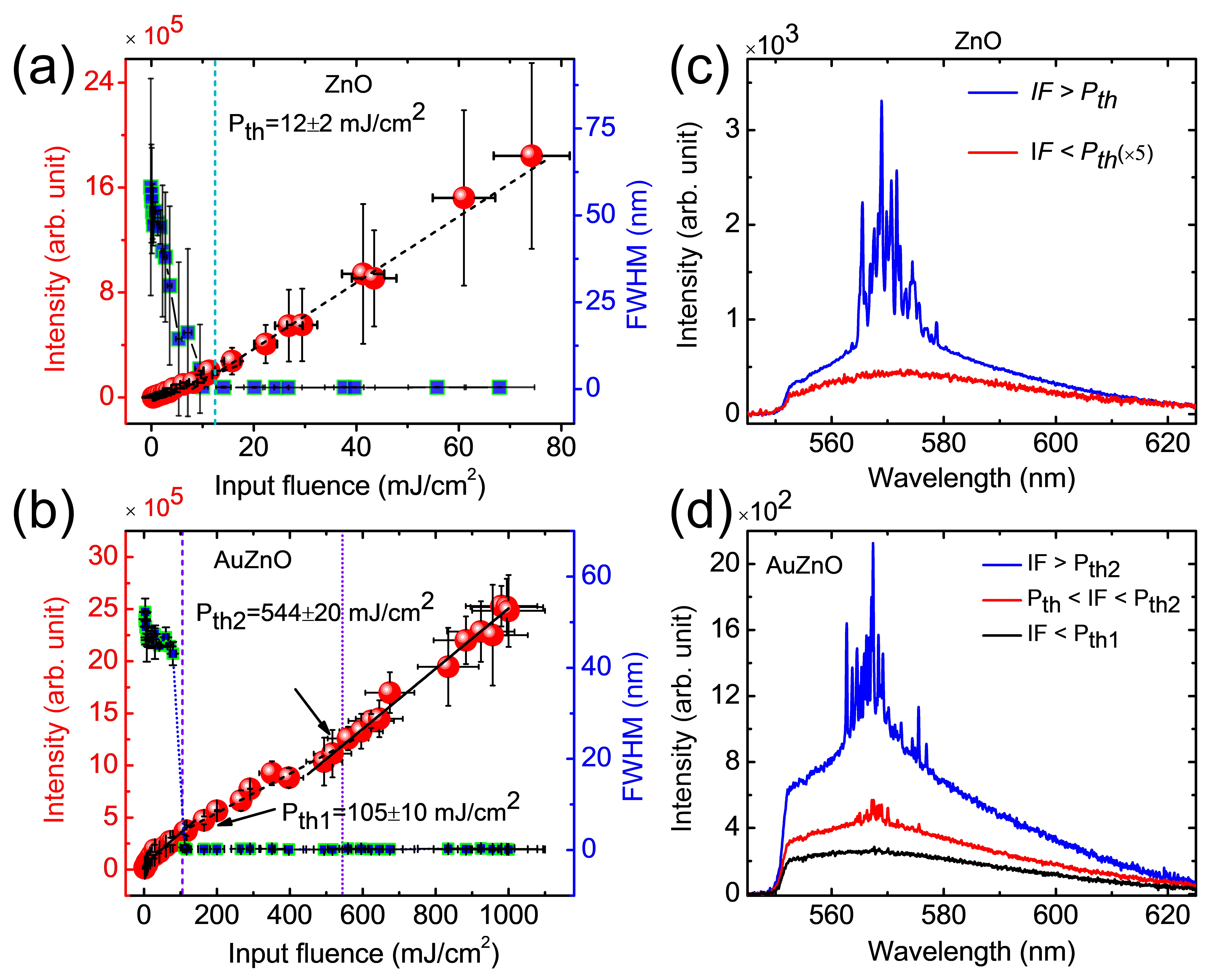}
		\caption{ Lasing threshold characteristics (a) and (b), and emission spectra at different input fluences (c) and (d) for ZnO and AuZnO, respectively. $P_{th}$ refers to the lasing threshold of the systems and $IF$ is the input pump fluence.}
		\label{fig37}
    \end{figure}
 
\subsection{Lasing Characteristics}

 Fig. \ref{fig37} (a) shows the lasing threshold characteristics of Pyrromethene-doped PMMA coated ZnO sample. A kink in the input fluence vs output intensity (I-O) plot with an increased lasing efficiency accompanied by the narrowing of the full width at half maximum (FWHM) of the emission spectra indicates the regular lasing threshold behaviour \cite{siegman1986lasers}. Above the lasing threshold multiple sharp spikes over the broad amplified spontaneous emission (ASE) pedestal in the emission spectrum indicate a multimode lasing behaviour as shown in Fig. \ref{fig37} (c).

 However, the I-O plot for AuZnO exhibits a change in slope at two input fluences ($P_{th1}$ and $P_{th2}$) as shown in Fig. \ref{fig37} (b). At the first kink, the lasing efficiency decreases, however, FWHM of the modes in the emission spectra reduces from 22 nm to $\sim$ 0.2 nm. The emission spectrum at this input fluence ($P_{th1}$) exhibits multiple peaks over a broad ASE pedestal as shown in Fig. \ref{fig37} (d) in red. At the input fluence corresponding to second kink ($P_{th2}$) in I-O curve the lasing efficiency increases, however no significant change in the FWHM is observed. The emission spectra above  $P_{th2}$ exhibits multiple spikes over a second dominant pedestal along with the existing ASE pedestal as shown in Fig. \ref{fig37} (d) in blue.
 
    \begin{figure}[ht]
		\centering
		\includegraphics[width=8 cm]{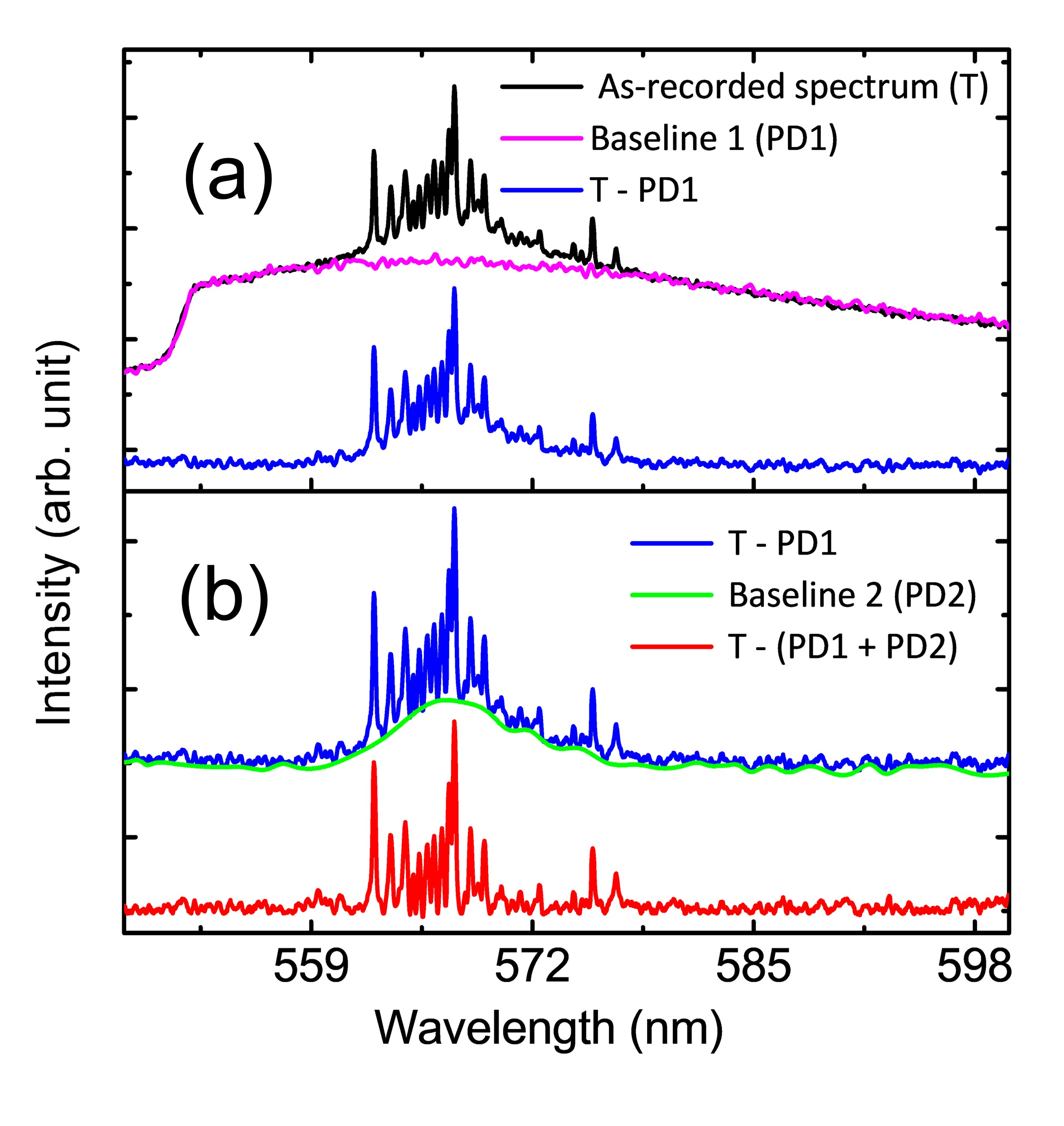}
		\caption{Separation of (a) baseline 1 (PD1) and (b) baseline 2 (PD2) from  the as-recorded spectrum to obtain coherent and incoherent contributions to RL emission spectra. The emission spectra, after pedestal removal, are offset along the the y-axis for better visualization.}
		\label{fig47}
    \end{figure}
 \subsection{Lasing Characteristics of Separated Out Spectral Components}

\begin{figure}[ht]
		\centering
		\includegraphics[width=8.5 cm]{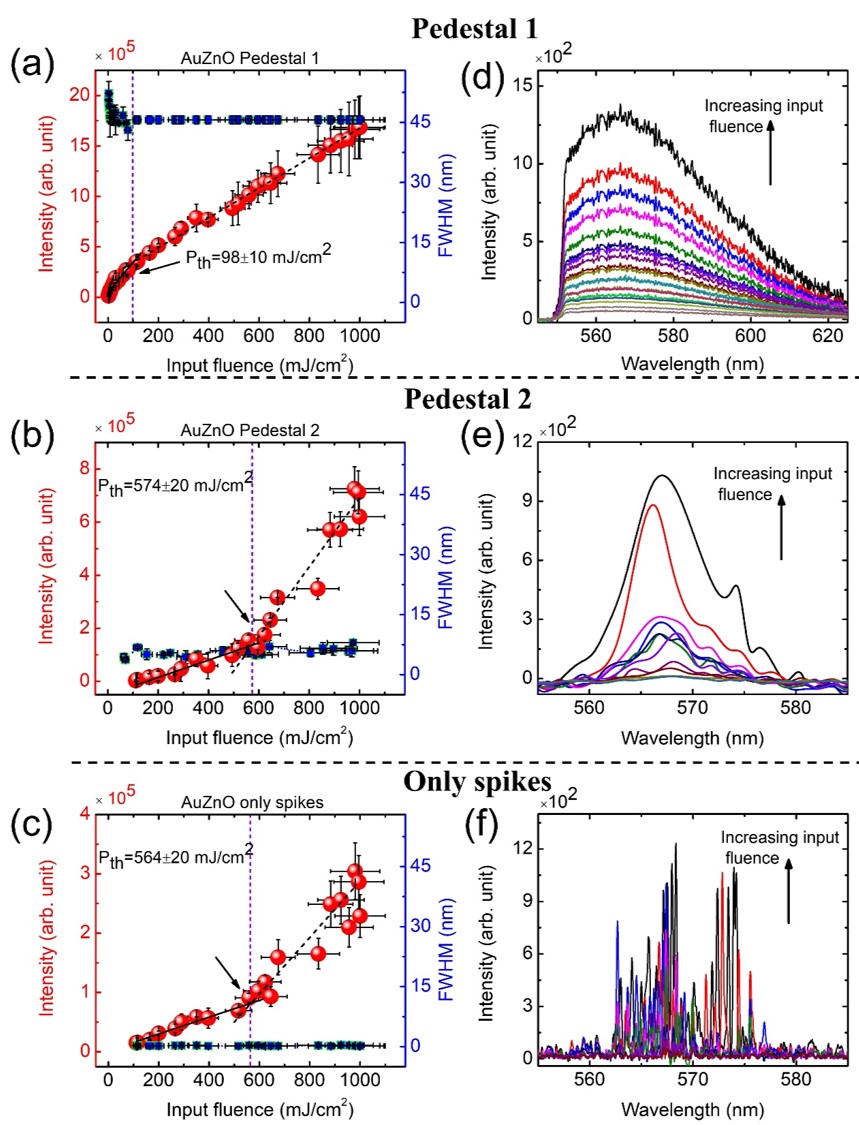}
		\caption{Threshold characteristics (a-c) and emission spectra (d-f) for the separated out pedestal 1, pedestal 2 and coherent spikes from the as recorded spectra, respectively.}
	\label{fig57}
      \end{figure}

 To understand the origin of the observed interesting double-threshold behavior in AuZnO RL, the incoherent and coherent contributions in the as-recorded RL emission (T)  were separated out by cubic spline interpolation of local minima of the emission spectra as shown in Fig. \ref{fig47} \cite{uppu2010statistical,uppu2011coherent,sarkar2021replica,yadav2023synergy}. Initially, the first pedestal (PD1) is separated out to obtain a spectrum consisting of second pedestal and coherent spikes as shown in Fig. \ref{fig47} (a). Then, the second pedestal (PD2) is   separated out too, to obtain only the coherent spikes in the spectrum as shown in Fig. \ref{fig47} (b). Thus the pedestals obtained correspond to incoherent emission and the coherent spikes in the spectra correspond to coherent emission. The variation of output intensity and FWHM of the emission spectra for the separated out components as a function of input fluence is shown in Figs. \ref{fig57} (a)-(c). It is observed that the I-O plot for PD1 exhibits a kink along with a reduction in FWHM of the emission by 4-5 nm  at the pump fluence close to $P_{th1}$ as shown in Fig. \ref{fig57} (a). Above $P_{th1}$ the slope for PD1 reduces similar to the behaviour of as-recorded spectrum in Fig. \ref{fig37} (b). However, no other kink is observed in the I-O plot for PD1 at higher pump fluences.
       \begin{figure*}[ht]
		\centering
		\includegraphics[width=17 cm]{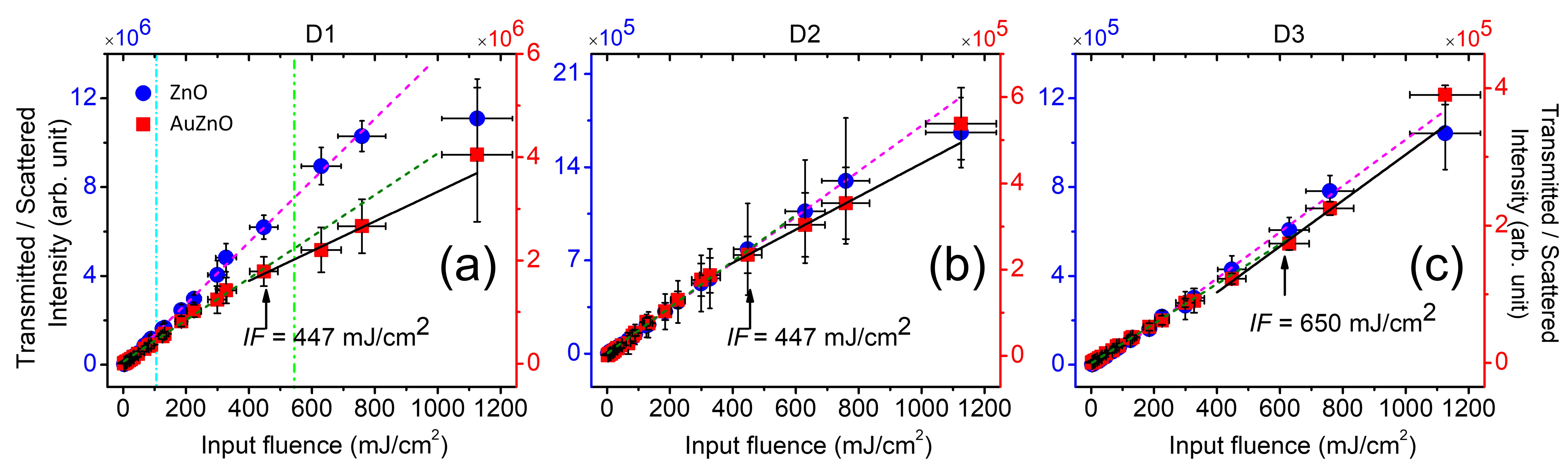}
		\caption{(a) Transmitted intensity including the scattered intensity (detector D1)  (b)  transmitted intensity with reduced scattered intensity contribution (detector D2) and (c) intensity scattered at an angle of 5 $^{\circ}$ from the beam propagation direction ( detector D3) from Pyrromethene-doped PMMA coated ZnO and AuZnO samples as a function of input fluence. The magenta and dark green lines are linear fits to the data. The cyan and light green lines indicate $P_{th1}$ and $P_{th2}$, respectively.}
	\label{fig67}
      \end{figure*} This decrease in slope of I-O plot for PD1 can be attributed to the clamping of the spontaneous emission as the system starts to lase \cite{kumar2021localized,santos2021gain}. However, a reduced lasing efficiency for the as-recorded spectrum indicates that even though the ASE gets clamped above $P_{th1}$, the stimulated emission does not build up at the same rate. This is only possible if there is an increased loss in the stimulated emission in the RL system above $P_{th1}$. The PD2 along with the coherent spikes start to appear above $P_{th1}$. Initially, the spikes are more dominant than PD2, but at higher pump fluences, PD2 becomes significant. Figs. \ref{fig57} (b) and (c) show the behaviour of output intensity and FWHM of the emission as a function of input fluence for PD2 and the coherent spikes in the emission, respectively. A regular lasing threshold type behavior is observed for both, with a threshold value close to $P_{th2}$. However, no change in FWHM of PD2 and spikes is observed above $P_{th2}$. It is to be noted that the change in the slope of I-O plot for PD2 occurs at slightly higher fluence as compared to the coherent part corresponding to only spikes in the emission. 
 The emission spectra for PD1, PD2 and spikes are shown in Figs. \ref{fig57} (d), (e) and (f), respectively, as a function of input fluence. Thus, the observed anomalous lasing behaviour in AuZnO RL can be attributed to an input fluence dependent interplay of absorption and gain in the system. A dominant absorption at $P_{th1}$ leads to a decreased lasing efficiency and at higher pump fluence a decrease in the absorption or increase in the effective gain in the system leads to the second threshold, $P_{th2}$.

\subsection{Nonlinear Scattering and Absorption Studies in AuZnO RL} \label{nls}
 The proposed pump fluence dependent changes in the absorption and gain were validated by probing the nonlinear optical absorption and scattering properties of the system. The nonlinear characteristics of a sample in general are studied using the standard Z-scan technique in which a laser beam is focused onto a sample using a lens. The sample is translated along the propagation direction of laser beam to vary the pump fluence \cite{sheik1990sensitive,venkatram2005nonlinear,kiran2004nonlinear,venkatram2006nonlinear}. In this study, the position of sample is kept fixed and the input fluence is varied using optical density filters. The optical setup in Fig. \ref{fig27} marked as NLO is used to probe the nonlinear optical behavior of AuZnO samples. The excitation setup for optical nonlinear studies is similar to that used for random lasing, but a different collection geometry has been employed. The transmitted and the scattered pump beam intensities from AuZnO sample were monitored as a function of pump fluence using three different configurations as shown in Fig. \ref{fig27} (labeled as NLO). The pump intensity transmitted from the AuZnO RL, collected using a configuration with f-number, $f/\#$ = 1 is recorded using a spectrometer referred to as D1. Before the light enters the collection fiber a ground glass diffuser (d) has been employed to get rid of the speckles in the transmitted light. The collection lens for D1 has large collection efficiency and is placed very close to AuZnO sample as shown in upper rectangle in Fig. \ref{fig27} (NLO). Thus, detector D1 collects the total transmitted light which includes  scattered light. Therefore, D1 only measures the losses due to the linear and nonlinear absorption of the sample. A configuration with $f/\#$ = 7.8 has been employed to collect transmitted light at a distance of $\sim$ 60 cm from the sample to reduce the scattered light collected by D2 as shown in lower rectangle in Fig. \ref{fig27} (NLO). As the distance from the sample increases the component of scattered intensity along the pump beam direction reduces. Hence, D2 collects predominantly the transmitted light from the sample and accounts for both absorptive and scattering losses in the system. A detector D3 employing a configuration with $f/\#$ = 7.8 has also been employed to collect light scattered at angle of 5 $^\circ$ from the pump beam propagation direction. Thus, D3 collects only scattered light and accounts for linear and nonlinear scattering from AuZnO sample. By monitoring intensities recorded by detectors, D1, D2 and D3, the nonlinear properties of the system can be deciphered.

Figs. \ref{fig67} show the pump intensities recorded in detectors D1, D2 and D3 as a function of input fluence for ZnO and AuZnO samples, respectively. A linear variation of the intensity recorded in D1, D2 and D3 with the pump fluence for ZnO RL indicates a linear behaviour of ZnO at the pump fluences of interest. For AuZnO RL, a linear variation of intensity recorded in D1 is observed for low pump fluences. As the pump fluence reaches $\sim$ 447  $\mathrm{mJ/cm^2}$ a nonlinear variation in the intensity collected by D1 is observed as shown in Fig. \ref{fig67} (a). The observed decrease in the transmitted intensity  indicates an enhanced absorption of pump beam in the system due to nonlinear absorption in AuZnO system at higher fluences. The transmitted intensity with reduced scattered intensity contribution, collected by D2 exhibits a similar variation as shown in Fig. \ref{fig67} (b). The decrease in intensity collected by D2 occurs at the same input fluence as for D1. This behaviour is expected as the transmitted intensity collected by D2 accounts for both optical absorption and scattering in the system. Hence, a nonlinear absorption observed by D1 will be reflected in D2 too. The variation of scattered light intensities recorded in detector D3 with the pump fluence is shown in Fig. \ref{fig67} (c). A linear behavior of scattered intensity is observed at low pump fluence, but, as the pump fluence reaches $\sim$ 650  $\mathrm{mJ/cm^2}$ an increase in the intensity collected by D3 is observed. Thus, an enhanced scattering from AuZnO RL at higher pump fluences is observed indicating nonlinear scattering in AuZnO system. However, the rate of increase in scattered intensity is low as compared to the rate of decrease in transmitted intensity. Hence, a distinct change in transmitted intensity collected by D2 is not observed at IF $\sim$ 650  $\mathrm{mJ/cm^2}$ despite an enhanced scattering. Thus, at higher input fluences the absorption and scattering of the pump beam increase nonlinearly in AuZnO RL. This nonlinear absorption and scattering of pump beam in AuZnO RL at high pump fluences lead to an enhanced gain and scattering feedback, respectively, leading to the occurrence of second lasing threshold, $P_{th2}$ in the system. The first lasing threshold, $P_{th1}$ can be attributed to the coupling of random lasing modes to the plasmonic nanocavities \cite{yadav2023synergy,yadav2023tracking}.At the lower pump fluence, the spontaneous emission in the system is homogeneously distributed in the system. As the lasing starts, ASE gets coupled to the modes of the system, leading to a redistribution of the emitted intensity into the  random lasing and plasmonic modes. The light confined in this cavities traverses these cavities several times before it leaks out. This results in an enhanced interaction between RL emission and the
plasmonic NPs. Depending on whether absorption or scattering by the plasmonic NPs is more dominant, a corresponding increase in quenching or enhancement of
the confined emission, compared to homogeneous spontaneous emission, is observed. For AuZnO sample a quenching of spontaneous emission observed in Fig. \ref{fig17} (d) indicates a dominant absorption of spontaneous emission by Au NIs. As the lasing starts, the coupling of RL modes to the plasmonic nanocavities leads to an enhanced absorption of RL emission as compared to ASE. Thus, even though the system has enough gain to lase, the increased absorption of lasing emission by plasmonic Au NIs above lasing threshold slows the output intensity's growth with input fluence. Hence, a coupling of RL modes to the plasmonic nanocavities consisting of Au NIs with dominant absorption leads to the unusual first lasing threshold of the system. In order to validate the observed double-threshold lasing behaviour, the fluctuation and correlations in the modes of AuZnO RL were studied using different statistical analysis tools.

\subsection{Fluctuations and Correlations of Modes of AuZnO RL}
\subsubsection{L\'evy Flight Statistics}
\begin{figure}[ht] 
		\centering
		\includegraphics[width= 8.5 cm]{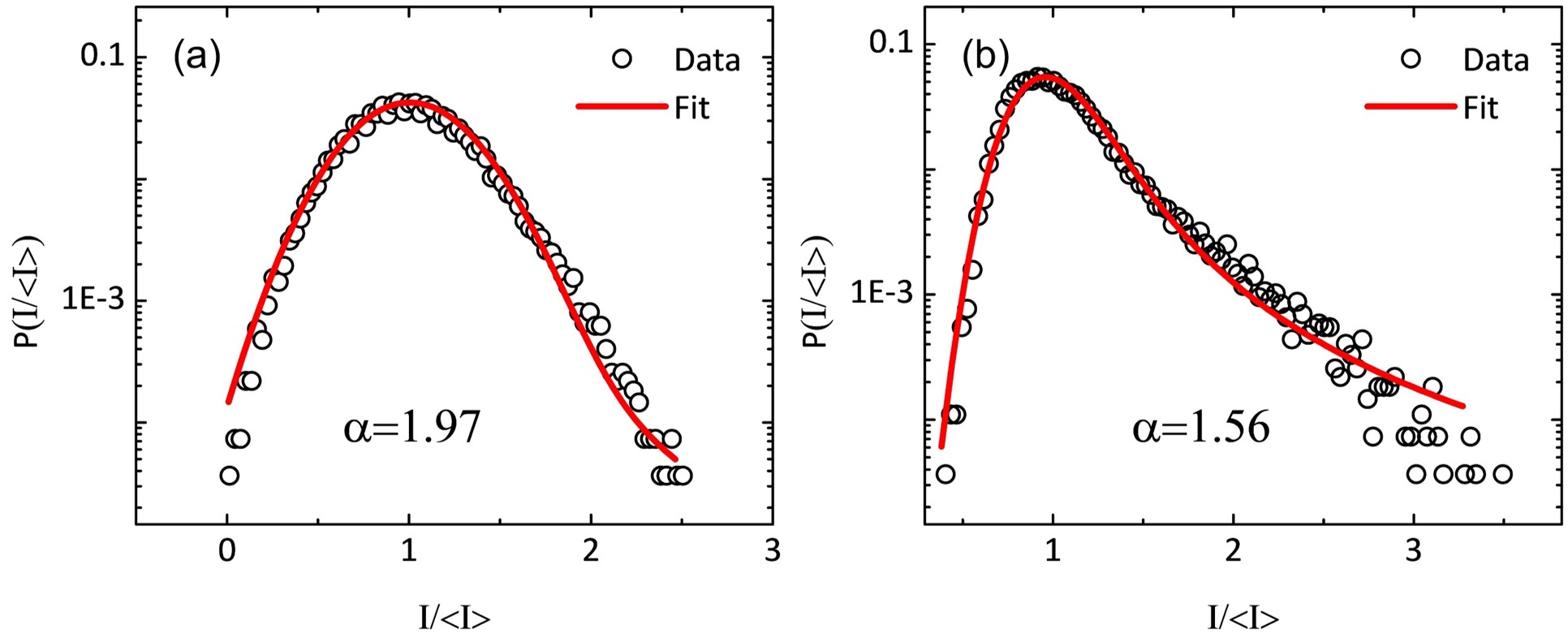}
		\caption{ The distribution of emission intensities for modes in 10 nm spectral range about gain maxima at (a)  $IF<P_{th1}$ and (b) $IF>P_{th1}$. The solid lines are the fit to the $\alpha$-stable distribution.}
		\label{alphafit}
    \end{figure}
 RLs exhibit intrinsic fluctuations in the lasing frequencies \cite{mujumdar2007chaotic}, the number of lasing modes \cite{wu2008statistical}, and emission intensities \cite{sharma2006levy,van2006intrinsic,lepri2007statistical,uppu2010statistical,uppu2012identification,zhu2012experimental,uppu2013dependence}. These fluctuations, despite the quenched disorder, are attributed to the presence of multiple modes with similar lasing thresholds and strong inter-modal interactions. \cite{mujumdar2007chaotic}. It has been observed that the emitted intensity distribution has features of L\'evy statistics, including a power-law decay \cite{wu2007statistics,lepri2007statistical,uppu2012identification,zhu2012experimental,uppu2013dependence,merrill2016fluctuations}. Depending on different parameters such as pump energy, system size, and disorder strength, the fluctuations may follow Lévy or Gaussian statistics. \cite{sharma2006levy,uppu2012identification,uppu2013dependence}. In order to characterize the statistical regime of RLs, the intensity distribution is fitted to an $\alpha$-stable distribution \cite{gnedenko1968limit}. The  $\alpha$-stable function describes a heavy-tailed distribution employing four parameters, the tail exponent, the location parameter, the skewness parameter and width parameter. The tail exponent, $\alpha \hspace{2pt}  \epsilon  \hspace{2pt}(0 \hspace{2pt} 2]$ describes the rate of tapering of the tail, with $\alpha=2$ indicating a Gaussian behaviour and for $\alpha < $  2 indicating a L$\acute{e}$vy behavior \cite{uppu2012identification,uppu2014levy}. The $\alpha$ parameter has been identified as a universal indicator of lasing threshold in RLs \cite{uppu2014levy}. To obtain $\alpha$, an iterative regression-based technique introduced by Koutrouvelis \cite{koutrouvelis1980regression} has been utilized, with initial estimates computed using the quantile-based McCulloch method \cite{mcculloch1986simple}. The intensities of modes in a spectral range of $\sim$ 10 nm around the emission maxima for 100 single shot spectra have been considered for $\alpha$-stable fitting. Figs. \ref{alphafit} (a) and (b) show the probability distribution of intensity of modes fitted to a $\alpha$-stable distribution at $IF<P_{th1}$ and $IF>P_{th1}$, respectively.
 \begin{figure}[t]
	\centering
		\includegraphics[width=8.2 cm]{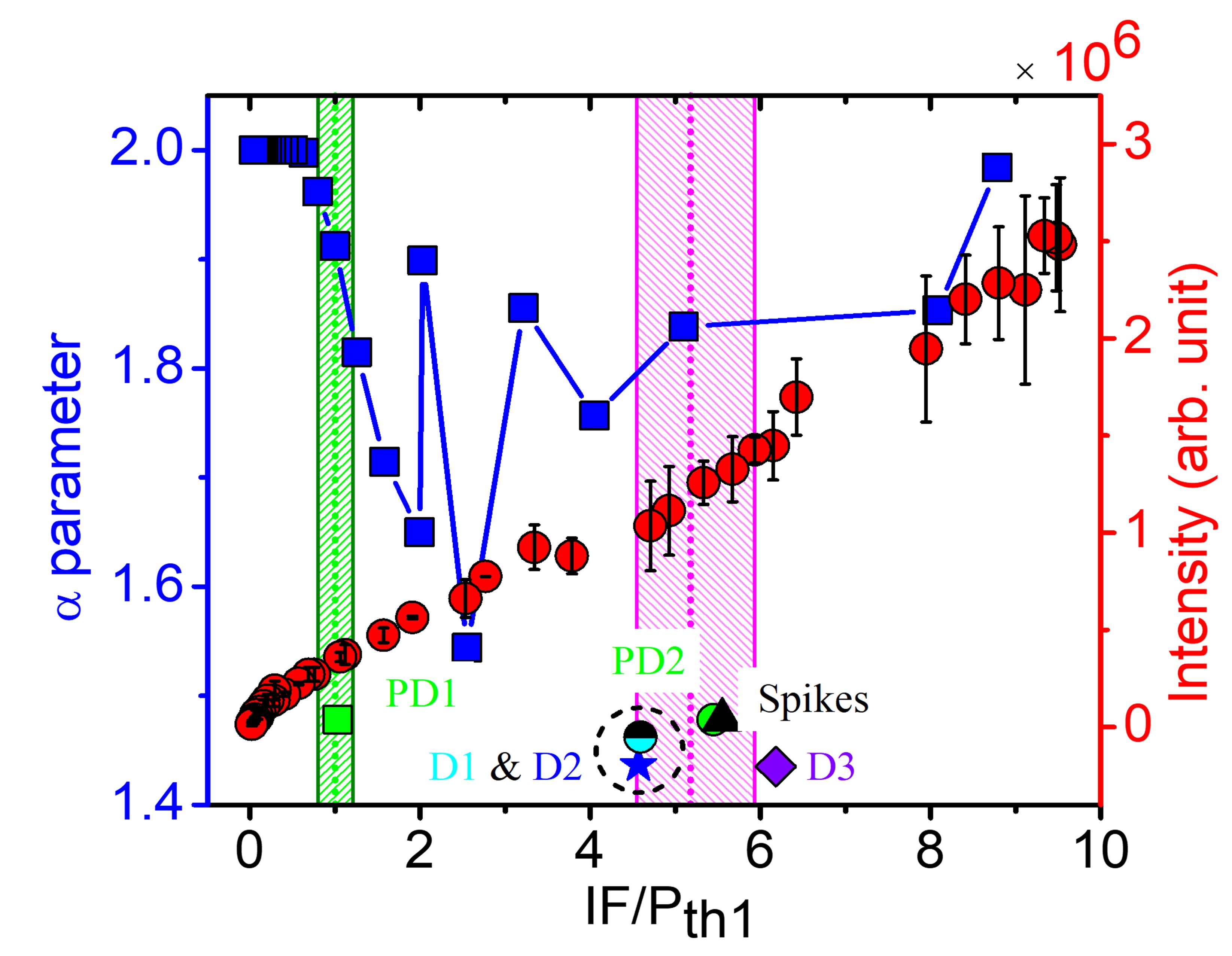}
		\caption{Variation of $\alpha$ parameter for modes in a spectral range of 10 nm around the emission maxima and total emission intensity of RL as a function of input fluence (here expressed in terms of ratio of input fluence and $P_{th1}$) for AuZnO RL. The dotted green and magenta lines represent $P_{th1}$ and $P_{th12}$, respectively. The shaded region around them represents a range of variation of lasing thresholds due to variation of gain across the sample and pump laser intensity fluctuations. Different individual points represent threshold of different entities as follows, green $\square$:   pedestal 1, green o: pedestal 2 , black $\triangle$: coherent spikes. The pump fluence corresponding to the emergence of nonlinear absorption and scattering in the system are indicated by, half cyan filled black o: D1, blue $\star$: D2 and violet $\diamond$: D3.}
		\label{fig77}
    \end{figure}
Fig. \ref{fig77}  shows the variation of $\alpha$ with the input fluence with respect to $P_{th1}$. In all the statistical analyses the pump fluence has been represented as a ratio of input fluence to $P_{th1}$ for the ease of comparison of statistical results with lasing thresholds of AuZnO RL system. Due to a finite number of measurements, an ambiguity in the regime of operation of RL arises for  the values of $\alpha$ close to 2. Hence $\alpha$ = 1.8 has been chosen as the threshold of transition from Gaussian to L$\acute{e}$vy regime \cite{ignesti2013experimental}.  It is observed that at low input fluence, $\alpha >$ 1.8 and at  $ IF= P_{th1}$, it falls below 1.8 indicating a transition from Gaussian to L\'evy regime. With further increase in pump fluence ($P_{th2} > IF > P_{th1}$) $\alpha$ fluctuates in the range [1.5, 1.8]. This fluctuating behavior can be attributed to large fluctuations in the system due to a competition between the effective gain and loss due to dominant absorption of pump and emission intensities, respectively, by Au NIs. Above $P_{th2}$, $\alpha $ again attains values $>$ 1.8.  The variation of the emission intensity of AuZnO RL with pump fluence is also shown for reference. The first and second lasing threshold of the system are marked with dotted green and magenta lines, respectively. The shaded regions around these lines correspond to the range of variation in the lasing threshold values due to fluctuation in pump laser intensity and variation of gain across different positions in the sample. The thresholds of separated out components, pedestal 1 ( green $\square$), pedestal 2 (green o ), and coherent spikes ( black $\triangle$) coincide well with range of $P_{th1}$ and $P_{th2}$. The pump fluence corresponding to the emergence of nonlinear absorption and scattering in the system are indicated by D1 (half cyan filled black o), D2 (blue $\star$) and D3 (violet $\diamond$). The onset of nonlinearity in the system occurs after $P_{th1}$ and before $P_{th2}$. Thus, a dominant absorption of RL emission due to coupling of RL modes to plasmonic Au nanocavities at the onset of lasing and a nonlinear absorption and scattering of the pump beam at the much higher input fluences leads to the anomalous lasing behaviour in the AuZnO RL.

\begin{figure*}[t] 
		\centering
		\includegraphics[width= 1\linewidth]{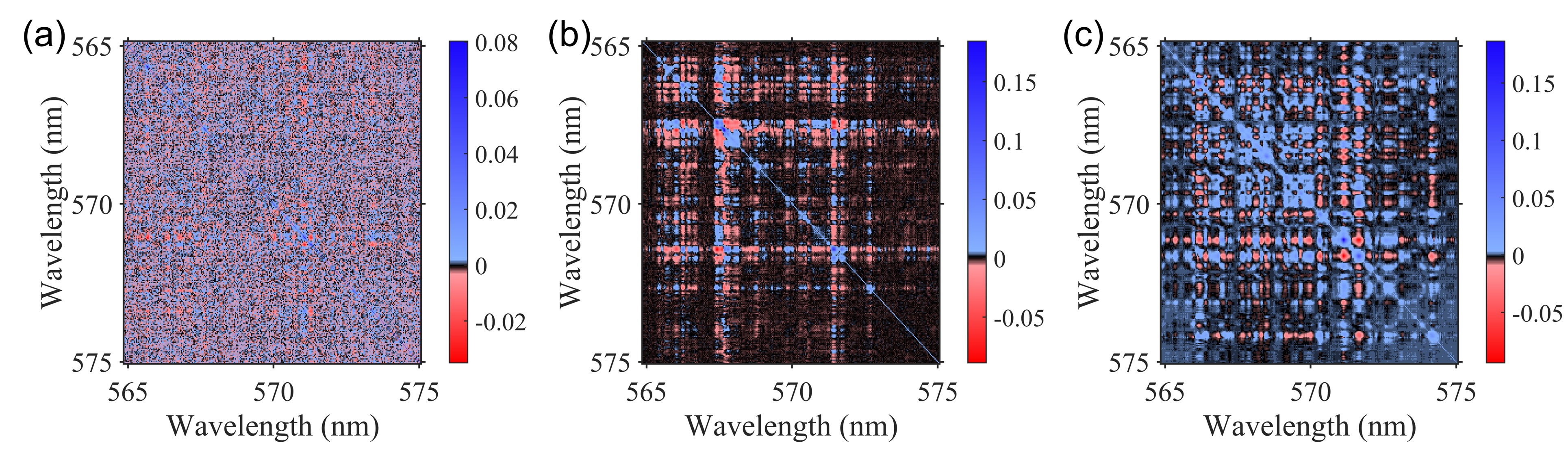}
		\caption{ Cross-correlation maps of RL modes at pump fluence (a) $IF<P_{th1}$ (b) $P_{th1}<IF<P_{th2}$, and (c) $IF>P_{th2}$.}
		\label{cova}
    \end{figure*}
\subsubsection{Covariance Analysis}
It has been observed that the intensity fluctuations in a RL are coupled across lasing modes due to inter modal interactions \cite{merrill2016fluctuations}. The correlation in the intensities of two lasing modes at $\lambda_1$ and $\lambda_2$ have been quantified using covariance \cite{merrill2016fluctuations,sarkar2020replica,choubey2020random}, 
\begin{equation}
    c_2 (\lambda_1,\lambda_2)=\langle I(\lambda_1) I(\lambda_2)\rangle_t-\langle I(\lambda_1) \rangle_t \langle I(\lambda_2) \rangle_t
\end{equation}

\noindent Prior to calculating the covariance, each spectrum's intensity is normalized by its mean intensity. Figs. \ref{cova} (a), (b) and (c) show the covariance maps at three different pump fluences, $IF<P_{th1}$, $P_{th1}<IF<P_{th2}$ and $IF>P_{th2}$, respectively. The diagonal elements in the covariance plots represent the auto-correlations of the modes. Below the lasing threshold in Fig. \ref{cova} (a), the cross-correlations are much smaller than the auto-correlations along with a very random distribution of positive and negative values indicating dominant noise and no mode coupling. Above $P_{th1}$ and $P_{th2}$, strong negative cross-correlations in red streaks separated by islands of blue, positive correlations are observed due a strong mode competition as shown in Figs. \ref{cova} (b) and (c). Thus, when the system starts to lase, the modes get coupled via the available gain in the system and compete with each other leading to a pool of large positive and negative correlations.

 \subsubsection{Replica Symmetry Breaking Analysis}
  \begin{figure}[ht] 
		\centering
		\includegraphics[width=8.5 cm]{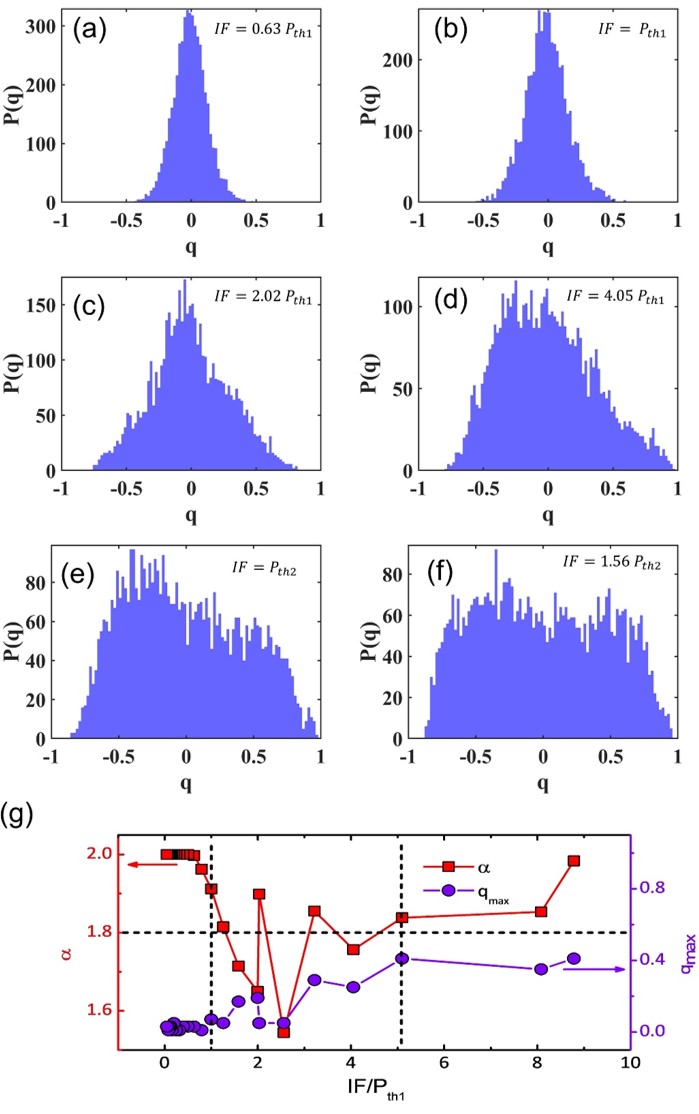}
		\caption{ (a)-(f) The distribution of Parisi overlap parameter at different pump fluences. (g) The variation of $q_{max}$ and $\alpha$ parameters as a function of pump fluence. The black dashed lines correspond to $P_{th1}$ and $P_{th2}$.}
		\label{rsb}
    \end{figure} 
 The shot-to-shot intensity fluctuations in RLs lead to a phenomenon called replica symmetry breaking (RSB) which was originally predicted by spin-glass theory \cite{mezard1987spin,angelani2006glassy}. According to spin-glass theory, the exact replicas of a system realized under identical experimental conditions can reach to different equilibrium states due to a large number of competing equilibrium states. The statistical distribution of an overlap parameter $q$, known as the Parisi overlap function $P(q)$, for such systems exhibits a change its shape. In RLs, each excitation laser pulse corresponds to a replica of the RL system. The lasing modes in RL emission are considered as the spin variable and the pump energy acts as the inverse temperature. At low pump energy, when the system is below the lasing threshold the RL modes do not interact with each other leading to a non-interacting paramagnetic regime. As the pump energy increases, the modes start to compete with each other for the limited gain due saturation of gain and hence, lasing modes oscillate in a correlated manner leading to a  correlated spin-glass phase. \cite{sarkar2020replica,choubey2020random}.
 The overlap parameter is defined as,
 \begin{equation}
     q_{\alpha \beta}= \frac{\sum_{\lambda=1}^{N} \Delta_{\alpha}(\lambda)\Delta_{\beta}(\lambda)}{\sqrt{\left[ \sum_{\lambda=1}^{N} \Delta_{\alpha}^{2}(\lambda)\right]\left[ \sum_{\lambda=1}^{N} \Delta_{\beta}^{2}(\lambda)\right]}}
 \end{equation}
 where, $\Delta_{\alpha}(\lambda)=I(\lambda)_{\alpha}-\bar{I}(\lambda)$ is the intensity fluctuation about the mean intensity $\bar{I}(\lambda)$ averaged over n pulses and $\alpha,\beta=1,2,...,n$ is the number of pulses. $N$ is the number of spectral points. The distribution of $q$ is evaluated as $P(q)$ and its shape determines the phase of the system. The value of $q$ at which $P(|q|)$ reaches the maximum is defined as $q_{max}$. In the photonic paramagnetic phase, $q_{max}=0$ while in the spin-glass phase $q_{max}\ne0$. 
 
 Fig. \ref{rsb} shows the $P(q)$ for different excitation energies. At the pump fluence below and near $P_{th1}$, $q_{max} \sim 0$ indicating that modes are not interacting, as shown in Figs. \ref{rsb} (a) and (b). In Figs. \ref{rsb} (c) and (d) it is observed that $q_{max} $ is still near zero, however a non-trivial distribution is observed due to broadening of $P(q)$. Even at higher input fluences further broadening of $P(q)$ is obtained without any appearance of  usual $q_{max}=1$ nature above lasing threshold as shown in Figs. \ref{rsb} (e) and (f). This behaviour has been observed in a few RL systems at very high pump energies after the system has reached spin-glass phase and is referred to as  unsaturated photonic spin-glass phase \cite{gomes2016observation,moura2020nonlinear}. Thus, in our case a transition from paramagnetic phase to unsaturated spin-glass phase is observed without any appearance of saturated spin-glass phase.  
 
 The variation of $q_{max}$  and $\alpha$ with  the pump fluence is shown in Fig. \ref{rsb} (g). It is observed that change in $\alpha$ and $q_{max}$ parameters coincides with the two lasing thresholds of AuZnO RL. However, below $P_{th1}$, $q_{max}\sim 0$, for $P_{th1}< IF < P_{th2}$, the $q_{max}$ value increases from 0 to 0.5 with certain fluctuations, and at $IF >P_{th2}$, the $q_{max}$ values remain constant at $\sim$ 0.5. Thus, the modes start to interact with each other after first lasing threshold, however they do not oscillate in completely correlated manner.

 
 \section{Conclusion}
 In conclusion, random lasing in a system with vertically aligned ZnO nanorods coated with a dye-doped polymer matrix has been studied. The effect of plasmonic Au NIs on the lasing characteristics of the system has been explored. The role of linear and nonlinear absorption and scattering properties of Au NIs has been deduced to understand the unique double-threshold behaviour of the system. The unusual first lasing threshold has been attributed to the dominant absorption of laser emission due to coupling of RL modes and plasmonic nanocavities. The second lasing threshold typically results from enhanced absorption and scattering of the pump beam due to the optical nonlinear behavior of Au NIs at high excitation energies. Several statistical analysis tools such as, L\'evy statistics, RSB and covariance analysis have been employed to validate the two lasing thresholds observed in AuZnO RL. Covariance analysis reveals that the modes start to interact with each other after the first lasing threshold despite the decreased lasing efficiency. The variation in the parameters, $\alpha$ and $q_{max}$ with pump fluence clearly indicates the presence of two distinct lasing thresholds in the system.
	
	\begin{acknowledgments}
		The authors acknowledge Jonathan Andreasen, Georgia Tech Research Institute for fruitful discussions. We acknowledge support from Science and Engineering Research Board (CRG/2020/002650) sponsored project. DST-FIST facility, Department of Physics, IIT Kharagpur is also acknowledged. 
	\end{acknowledgments}
     \appendix
%

	\end{document}